\documentclass[twocolumn,aps,pra,superscriptaddress,longbibliography]{revtex4-2}

\usepackage{amsmath,amssymb,graphicx,hyperref,bm}
\usepackage{hyperref}

\usepackage{booktabs} 

\usepackage{xcolor} 
\usepackage{colortbl} 
\usepackage{orcidlink} 
\usepackage{subcaption} 
\newtheorem{definition}{Definition}

\newcommand{\figref}[1]{Figure~\ref{#1}}

\begin{document}

\title{Local surrogates for quantum machine learning}

\author{Sreeraj Rajindran Nair \orcidlink{0009-0003-8753-5621}}
\affiliation{Centre for Quantum Software and Information, University of Technology Sydney, Australia}

\author{Christopher Ferrie \orcidlink{0000-0003-2736-9943}}
\affiliation{Centre for Quantum Software and Information, University of Technology Sydney, Australia}

\begin{abstract}
Surrogates have been proposed as classical simulations of the pretrained quantum learning models, which are capable of mimicking the input-output relation inherent in the quantum model.  Quantum hardware within this framework is used for training and for generating the classical surrogates. Inference is relegated to the classical surrogate, hence alleviating the extra quantum computational cost once training is done. Taking inspiration from interpretable models, we introduce a local surrogation protocol based on reuploading-type quantum learning models, including local quantum surrogates as cost-efficient intermediate quantum learning models. When the training and inference are only concerned with a subregion of the data space, deploying a local quantum surrogate offers qubit cost reductions and the downstream local classical surrogate achieves dequantization of the inference phase. Several numerical experiments are presented.

\end{abstract}

\maketitle


\section{Introduction}

A widely employed class of quantum learning models uses a parameterized quantum circuit to encode input data and generate predictions. These models are often referred to as quantum neural networks \cite{McClean_2016, PhysRevA.98.032309, dunjko2017machinelearningartificial, Wang_2024}, as the parameters of the circuit loosely resemble the weights in traditional neural networks. 

Reuploading quantum models \cite{P_rez_Salinas_2020, P_rez_Salinas_2021} fall under this quantum neural networks class, structured with alternating layers of data-encoding gates and trainable unitary gates. The input data is reuploaded multiple times throughout the circuit. It has been shown that these reuploading models can be exactly represented as truncated Fourier series \cite{PhysRevA.103.032430}. Utilizing this unique property of such models, Schreiber et al \cite{Schreiber_2023} proved that re-uploading models always warranted Fourier-based classical surrogates. With this, the inference phase of such quantum learning models was de-quantized \cite{cotler2021revisitingdequantizationquantumadvantage, cerezo2024doesprovableabsencebarren}, as the quantum hardware is only required for the training phase and for the subsequent creation of the classical surrogate. 

Jerbi et al \cite{Jerbi_2024} further expanded upon this concept beyond just the reuploading models to propose a general framework for 'shadow models' and conditions for 'shadowfiable' quantum learning models. Shadows and surrogates are interchangeable terms within our context \cite{basheer2022alternatinglayeredvariationalquantum}. In addition to classical simulations \cite{rudolph2023classicalsurrogatesimulationquantum, cerezo2024doesprovableabsencebarren, Shao_2024}, surrogates taking inspiration from classical machine learning \cite{hong2021surrogatebasedsimulationoptimization} have been used for optimization purposes as well \cite{PhysRevA.107.032415}. Moreover, quantum surrogates for quantum models are also an active area of research \cite{nakayama2024explicitquantumsurrogatesquantum}. Throughout the rest of this paper, quantum models/surrogates refer to learning models that require quantum hardware in the inference phase, and classical models/surrogates refer to models where classical hardware would suffice for the same.

The aforementioned works were concerned primarily with the surrogation of the whole data space of the trained quantum model. Taking a cue from classical machine learning interpretability techniques \cite{molnar2025}, we refer to such surrogation protocols as {\bf global surrogation} in this paper.

\begin{figure}[t]             
  \centering
  \begin{subfigure}[b]{\columnwidth}
    \centering
    \includegraphics[width=\linewidth]{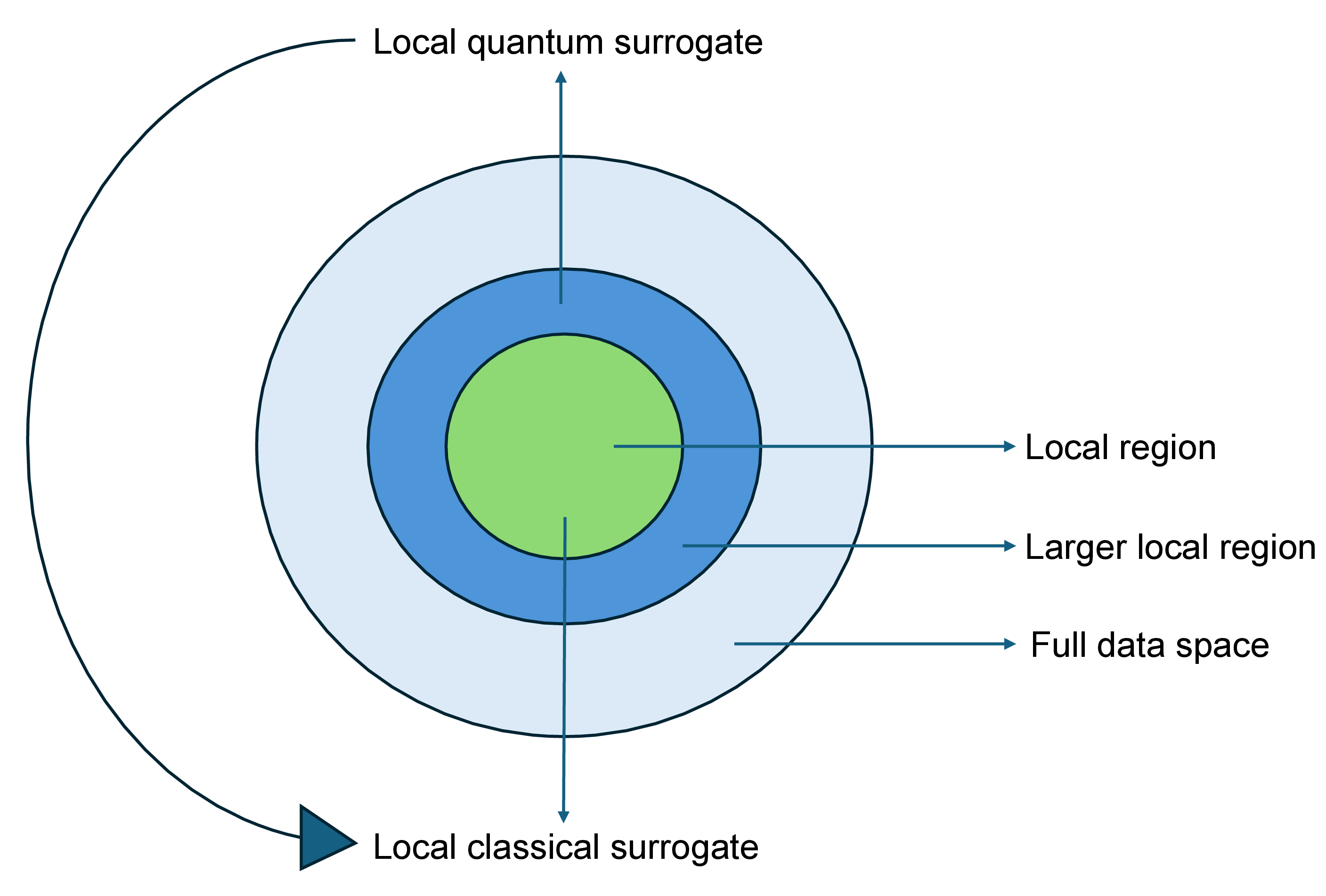}
  \end{subfigure}
  \caption{Local surrogation. For a small enough local region within the data space of a quantum learning model, a cost-efficient local quantum/classical surrogate can be generated with adequate accuracy, potentially as a two-step protocol.   }
  \label{fig:qubit_1d}
\end{figure}

Our work focuses on surrogation when performed on smaller regions within the fully trained arbitrary quantum model data space, dubbed {\bf local surrogation}. We consider both local \emph{quantum} surrogates and local \emph{classical} surrogates. We numerically demonstrate that any arbitrary quantum model can be locally quantum surrogated with a reuploading-type model up to a certain error threshold within a local region of the data space. Once such a quantum Fourier local surrogate is generated, it is always possible to construct a local classical surrogate using the classical surrogation protocol proposed by Schreiber et al \cite{Schreiber_2023}. Within our framework, the sizes of local regions for the local quantum surrogate and the local classical surrogate are mutually independent. Although one can apply a modified version of the classical-surrogation protocol directly on the arbitrary quantum learning model, the global spectrum typically grows combinatorially with circuit depth and data dimension, and no tight cost guarantees currently exist. 

Local surrogates are used extensively in classical machine learning within the context of interpretable/explainable AI (XAI) \cite{Longo_2024}. For instance, Local Interpretable Model-agnostic Explanations (LIME) \cite{10.1145/2939672.2939778} is an approach to explain the individual predictions of a black-box machine learning model. Recently, this framework was extended to quantum machine learning as quantum local interpretable model-agnostic explanations (QLIME) \cite{Pira_2024}. Likewise, localization in the data space is a staple of digital signal processing (DSP) \cite{10.5555/281875, oppenheim1999discrete}. In DSP, this usually refers to the use of windowing techniques \cite{prabhu} when processing large or real-time signals, as seen in short-time Fourier transform (STFT) processing \cite{allen1977short, allenrabiner1977, DUHAMEL1990259, stockham}. Windowing broadly refers to the operation of chopping up signals/data into smaller chunks and processing these with standard window functions \cite{bergold2020fourierserieswindowedbump}. Taking a cue from DSP, we refer to the local regions in the data space as windows and the size of these local regions as window sizes later in our simulations. The terminology is appropriate, given that both formalisms share a common Fourier representation.


A recent paper \cite{lerch2024efficient} introduced a hybrid quantum-classical simulation framework designed to approximate expectation landscapes generated by quantum circuits. They focus on small subregions or ``patches'' of these landscapes, defined over a subregion, using a classical surrogate. The work purports that it is indeed possible to efficiently simulate the quantum system classically with polynomial complexity and to efficiently distribute computational labour between quantum and classical hardware by essentially minimizing the use of quantum hardware to the absolute bare minimum. The conceptualization of these ``patches'' is similar to the local region or windows from our framework. However, in contrast to our work, they do not delve into the windowing techniques \cite{prabhu} from digital signal processing \cite{10.5555/281875, oppenheim1999discrete} nor are they concerned with localization in the data space as we are.

The structure of the paper follows. We begin by discussing reuploading quantum models and their Fourier representation. We reiterate its generalization to the multidimensional case and follow it up with its expressivity. Then we review the classical surrogation protocol proposed by Schreiber et al, along with their definitions for classical surrogates. With these background materials covered, we introduce our local surrogation protocol and its potential advantages. We discuss the details and results of our simulations to validate our claims before giving a final conclusion at the end.

\section{Reuploading Quantum Models}

A reuploading quantum model consists of alternating layers of data encoding gates \( S(x) \) and trainable unitaries \( W(\theta) \) \cite{P_rez_Salinas_2020, P_rez_Salinas_2021}. The model's output at a specific input \( x \) is given by the expectation value of an observable \( M \) with respect to the quantum state prepared by the circuit:
\begin{equation}
f_\theta(x) = \langle 0 | U^\dagger(x, \theta) M U(x, \theta) | 0 \rangle,
\end{equation}
where the circuit \( U(x, \theta) \) has the structure:
\begin{equation}
U(x, \theta) = W^{(L+1)}(\theta) S(x) W^{(L)} (\theta)\cdots S(x) W^{(1)}(\theta).
\end{equation}
Here, \( L \) is the number of reuploading layers, \( S(x) = e^{-i x H} \) encodes the input \( x \) into the quantum state using a Hamiltonian \( H \), and \( W^{(k)} (\theta)\) are the trainable unitaries parametrized by \( \theta \).

\subsection{Fourier Representation}
Earlier work has proven that the natural representation of \( f_\theta(x) \) is as a truncated Fourier series \cite{PhysRevA.103.032430, vidal2020inputredundancyparameterizedquantum, landman2022classicallyapproximatingvariationalquantum}. The data encoding gates \( S(x) \) introduce a frequency spectrum \( \Omega \) determined by the eigenvalues of the Hamiltonian \( H \). The overall model can be expressed as:
\begin{equation}
f_\theta(x) = \sum_{\omega \in \Omega} c_\omega(\theta) e^{i \omega x},
\end{equation}
where \( c_\omega(\theta) \) are the Fourier coefficients, determined by the trainable gates \( W(\theta) \) and the measurement observable \( M \).

\subsection{Truncated Fourier Series}
The frequency spectrum \( \Omega \) accessible to the model depends on the eigenvalues \( \lambda_i \) of \( H \). Specifically, the spectrum is given by:
\begin{equation}
\Omega = \{ \lambda_i - \lambda_j \, | \, i, j \in [d] \},
\end{equation}
where \( d \) is the dimension of the Hilbert space. The Fourier series is truncated in the sense that only frequencies in \( \Omega \) contribute to the model's output. The coefficients \( c_\omega(\theta) \) determine how these frequencies combine to form the final function \( f_\theta(x) \) \cite{carleson1966convergence}.


\section{Multidimensional Fourier Series}

A multidimensional Fourier series generalizes the concept of a one-dimensional Fourier series to higher dimensions. For an $M$-dimensional input $\bm{x} = (x_1, x_2, \dots, x_M) \in \mathbb{R}^M$, the series is represented as
\begin{equation}
    f(\bm{x}) = \sum_{\bm{\omega} \in \Omega} c_{\bm{\omega}} e^{i \bm{\omega} \cdot \bm{x}},
\end{equation}
where $\bm{\omega} = (\omega_1, \omega_2, \dots, \omega_M)$ are the frequency components and $c_{\bm{\omega}}$ are the Fourier coefficients. The frequency spectrum $\Omega$ contains all possible combinations of the frequency components up to a given degree $D$, ensuring that the series captures the desired expressivity.

The number of independent Fourier coefficients is given by
\begin{equation}
    N_c = \frac{(2D + 1)^M - 1}{2} + 1,
\end{equation}
which scales exponentially with the dimension $M$. This highlights the computational challenges associated with high-dimensional Fourier representations.

To implement multidimensional Fourier series using quantum circuits, earlier works \cite{PhysRevA.107.062612} considered different ansatzes that encode data into quantum states and apply parameterized gates for processing. Each ansatz balances expressivity and scalability differently. For our simulations, we used the so called ``line ansatz''. The line ansatz encodes all data dimensions into a single qubit sequentially. The encoding gate $S(x_m)$ acts on the $m$-th data feature, followed by a processing gate $A(\bm{\theta})$ that applies trainable parameters. The output is a Fourier series in which all dimensions are mixed, achieving expressivity at the cost of reduced scalability.

Note that the particular ansatz chosen to arrive at the Fourier representation is not a relevant parameter in this work, though in practice it may be an important application-dependent factor.

\section{Expressivity of Reuploading Models}
The expressivity of a quantum model is governed by two factors:
\begin{itemize}
    \item \textbf{Frequency Spectrum}: The set of accessible frequencies \( \Omega \) determines the model's ability to represent functions with varying oscillatory behavior. Increasing the number of layers \( L \) or using a richer encoding Hamiltonian \( H \) expands \( \Omega \).
    \item \textbf{Fourier Coefficients}: The trainable gates \( W(\theta) \) control the coefficients \( c_\omega(\theta) \), which determine how the accessible frequencies combine to approximate the target function.
\end{itemize}

For a single-layer model (\( L = 1 \)) with a Pauli-\( X \) encoding (\( H = \sigma_x / 2 \)), the spectrum consists of frequencies \( \{-1, 0, 1\} \). By repeating the encoding layer multiple times (\( L > 1 \)), the spectrum is extended to \( \{-L, -L+1, \dots, L\} \), allowing the model to represent higher-degree Fourier series.

The expressivity of quantum circuits for multidimensional Fourier series depends on the encoding strategy, the number of layers $L$, and the structure of the processing gates. The degrees of freedom increase with $L$, enabling the representation of higher-order frequencies. However, practical considerations such as gate fidelity and noise impose constraints on the achievable expressivity.

The scaling of the Fourier series coefficients is also influenced by the eigenvalues of the encoding Hamiltonian $H$. By optimizing the eigenvalue spectrum, we can enhance the coverage of the frequency spectrum and tailor the model to specific applications.

Reuploading quantum models can approximate any square-integrable function \( f(x) \) on a compact domain by increasing the number of layers \( L \) and using sufficiently flexible trainable gates \( W(\theta) \) \cite{PhysRevA.103.032430, vidal2020inputredundancyparameterizedquantum}. The universal approximation theorem for quantum models follows from the fact that Fourier series with arbitrarily many frequencies can approximate any \( L^2 \)-function to arbitrary accuracy \cite{carleson1966convergence}.


\section{Classical surrogation protocol}

Quantum machine learning (QML) explores the use of quantum circuits to approximate complex functions. Despite the potential advantages of quantum learning models, their deployment is limited by the constraints of NISQ devices \cite{Schuld_2022}. Although even with adequate fault-tolerant quantum infrastructure, it is unlikely that it would be possible or even recommended to deploy quantum models for all and every learning task due to the associated high computational cost. The notion of \textit{classical surrogates} provides a practical means to overcome this limitation by enabling efficient evaluation and inference of quantum models in classical settings \cite{Schreiber_2023, Jerbi_2024}.

In this section, we discuss how reuploading-type quantum models admit classical surrogation through their representation as truncated Fourier series \cite{Schreiber_2023, landman2022classicallyapproximatingvariationalquantum}. We also outline the classical surrogation protocol and its theoretical guarantees. We will now reiterate the definition for a classical surrogate as provided by Schreiber et al. for completion.

\begin{definition}[Classical Surrogate \cite{Schreiber_2023}]
A hypothesis class of quantum learning models $\mathcal{F}$ has classical surrogates if there exists a process $\mathcal{S}$ that upon input of a learning model $f \in \mathcal{F}$ produces a classical model $g \in \mathcal{G}$ such that the maximal deviation of the surrogate from the quantum learning model is bounded with high probability. Formally, we require:
\[
\mathbb{P} \left[ \sup_{x \in \mathcal{X}} \| f(x) - g(x) \| \leq \epsilon \right] \geq 1 - \delta,
\]
for a suitable norm on the output space $\mathcal{Y}$. The surrogation process $\mathcal{S}$ must be efficient in the size of the quantum learning model, the error bound $\epsilon$, and the failure probability $\delta$.
\end{definition}

Given the Fourier representation of a quantum model, a classical surrogate can be constructed by estimating the Fourier coefficients $\{c_\omega\}$. The protocol proceeds as follows:
\begin{enumerate}
    \item \textbf{Frequency Spectrum Identification:} Determine the accessible frequency spectrum $\Omega$ based on the structure of the encoding Hamiltonian $H$ and the number of data encoding layers $L$.
    \item \textbf{Sampling and Coefficient Estimation:} Evaluate the quantum model $f_\theta(x)$ at discrete points $\{x_j\}_{j=1}^T$ wherein \(T=(2L+1)^d\) (L = number of reuploading layers, d = number of data features) and solve the least-squares problem:
    \begin{equation}
        \hat{c}_\omega = \arg \min_{c_\omega} \sum_{j=1}^T \left| \sum_{\omega \in \Omega} c_\omega e^{i \omega x_j} - f_\theta(x_j) \right|^2.
    \end{equation}
    \item \textbf{Classical Surrogate Construction:} Construct the classical surrogate $g(x)$ as:
    \begin{equation}
        g(x) = \sum_{\omega \in \Omega} \hat{c}_\omega e^{i \omega x}.
    \end{equation}
\end{enumerate}






This approach ensures that the classical surrogate $g(x)$ reproduces the output of the quantum model $f_\theta(x)$ with high fidelity. The classical surrogate $g_c$, generated with the surrogation protocol, satisfies the surrogation conditions of Definition 1 with the following number of invocations of the quantum learning model:
\begin{equation}
    N_{\text{total}} = T N = \frac{2 T \| M \|^{2}_{\infty}}{\epsilon^2} \left( \log \frac{1}{\delta} + T \log 2 \right),
\end{equation}
where $\| M \|$ is the bound on the norm of the measurement operator applied to the quantum state. It characterizes the measurement's sensitivity.

It is evident that with a larger number of data features d, the classical surrogation protocol will become computationally intractable due to the exponential scaling on d.




\section{Local surrogation protocol}
The surrogation protocol described by Schreiber et al generates a {\bf global} surrogate for the reuploading-type quantum learning model. This broadly means that the surrogate tries to replicate the predictions of the quantum model for all of the input data space. However, there are cases where the intent behind the creation of a surrogate might not require such full-scale prediction replication. In cases where the intention is to generate a surrogate only for a local region within the data space of an arbitrary quantum learning model, we can deploy the {\bf local} surrogation protocol. We numerically show that for most applications, this is a sufficient model and results in a generous reduction in qubit resources. 

In this manner, our work aligns with the \cite{lerch2024efficient} wherein classical surrogates were created for smaller regions of the \emph{parameter} space. Whereas, our work concerns local surrogates in the data space and extends the surrogation from local classical surrogates to local quantum surrogates as well, which we introduce in this paper.

Given a fully trained arbitrary quantum learning model, we can create a local Fourier bases quantum surrogate with the reuploading architecture, which would replicate the predictions of the quantum learning model within the local region the quantum model was trained on. We train the reuploading-type local quantum surrogate on the inputs and outputs of the arbitrary quantum model within the local region. Because it is confined to a restricted input region, the local quantum surrogate can be trained with markedly fewer quantum-circuit evaluations and a lower-dimension least-squares fit than its global counterpart. 

In addition to being more compact and resource-efficient, a local quantum surrogate can directly seed the Fourier classical-surrogation protocol, which guarantees the construction of a corresponding local classical surrogate. Local quantum surrogates can act as a cost-efficient intermediate model for classical surrogate creation. 

While it is possible to perform a modified version of the classical surrogation protocol directly on the said arbitrary quantum model, the number of required query points and the size of the resulting linear system scale with the overall input space and circuit depth. Without any proven upper bound, runtime and memory requirements can quickly become prohibitive, and the fit may suffer from poor numerical conditioning for higher-dimensional data.




\section{Simulations}

We ran several numerical simulations to validate our local surrogation protocol. In the following simulation cases, we first train our local quantum and classical surrogates on sampled target functions; that is, \emph{data} will refer to the output of a previously trained quantum model. Later on, we ran the same simulations on a quantum support vector machine model. As before, here the window size refers to the size of the local region. We also tried to restrict ourselves to the least capable surrogate models, or rather low qubit count shallow circuits in each case to showcase that local surrogation can aid in leveraging and extracting better performance from hitherto lackluster model configurations.

\subsection{Local quantum surrogation on 1D target function}

\begin{figure}[t]              
  \centering
  \begin{subfigure}[b]{\columnwidth}
    \centering
    \includegraphics[width=\linewidth]{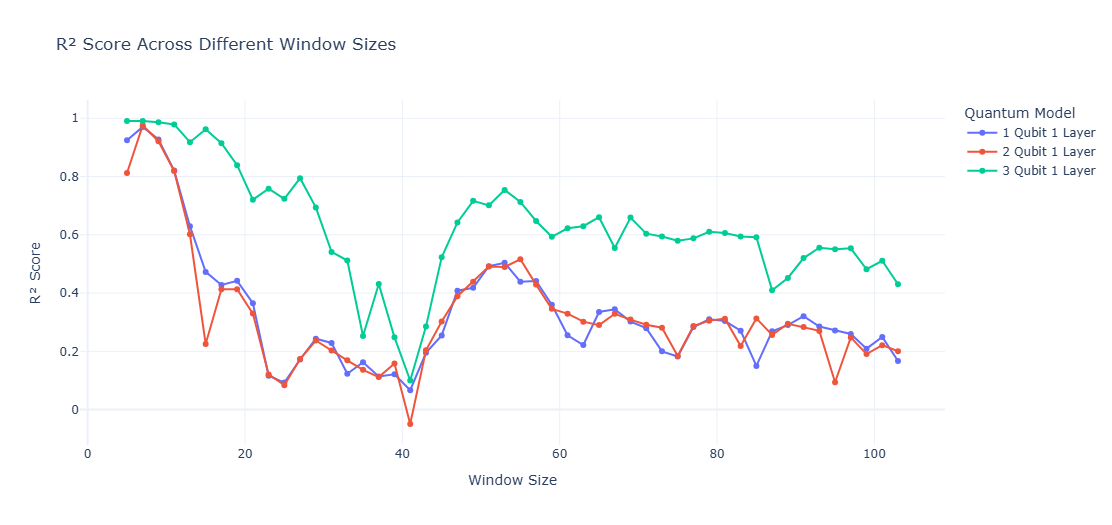}
  \end{subfigure}
  \caption{Testing 1D signal fitting in terms of improvement in $R^2$ learning metric in the case of 1, 2, and 3 qubit single-layer quantum reuploading models. The 1D signal under consideration is a truncated Fourier series of degree 3, i.e., $g(x) = \sum_{n=-5}^{5}c_{n}e^{2inx}$ with $c_{n} = 0.15 + 0.15i$ for $n=1,2,3$ and $c_{0}=0.1$. For the sampling rate of 10 within x values (-6, 6), the window begins as a 0.5-unit slice (i.e., 5 data points) of the data space, grows by 0.2 units every iteration, and maxes out before 120 data points.}
  \label{fig:qubit_1d}
\end{figure}

To test the effect of window size against learning metrics in the case of 1D local quantum surrogation, we ran simulations with a 1D target function directly. For such a target function, we train local surrogates for various window sizes. We tested the 1D target function fitting for improvement in $R^2$ learning metrics for the local region in the case of 1, 2, and 3 qubit single-layer quantum reuploading models using a modified version of the model described in \cite{PhysRevA.103.032430}. These models were trained with a maximum of 60 steps of an Adam optimizer with a learning rate of 0.3 and a batch size of 25. The target function was sampled at the rate of 10 samples per unit within the $x$ values $(-6, 6)$ for creating a net total of 120 data points. The window begins as a 0.5-unit slice (5 data points) of the data space, grows by 0.2 units every iteration, and maxes out at 120 data points. We see a clear trend of reduction in $R^2$ for the local region as the window size increases \figref{fig:qubit_1d}. The details of the target function are also provided in the \figref{fig:qubit_1d}.

\subsection{Local surrogation on 2D target functions}

\begin{figure*}[ht]
    \centering
    \begin{subfigure}[b]{\textwidth}
        \centering
        \includegraphics[width=\textwidth]{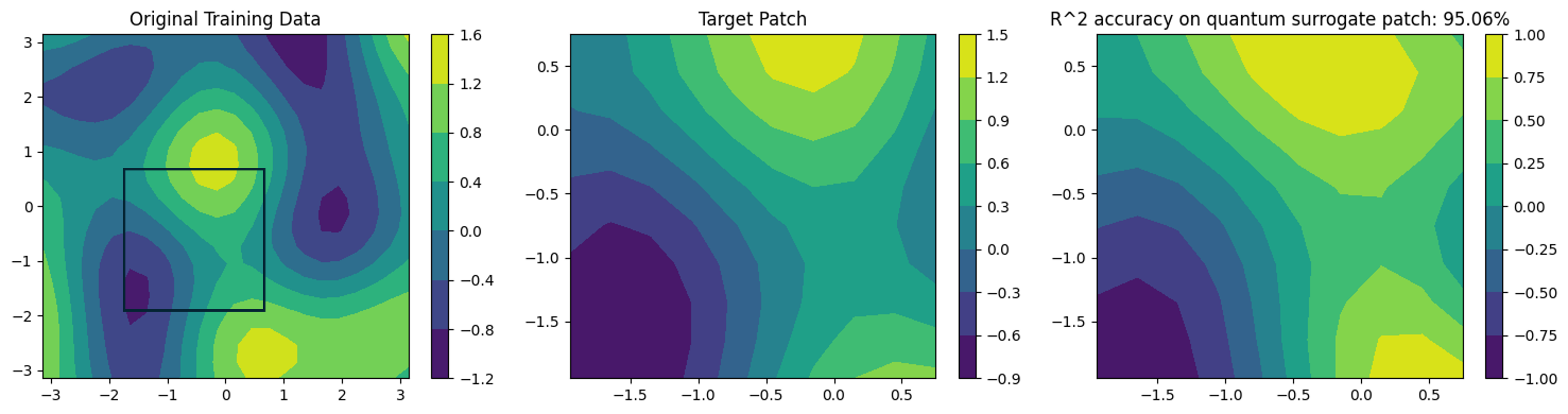}
        \caption{Local quantum surrogation of the target patch sampled from the target function $0.05x^{2} - 0.1y + 0.6cos(1.2x + 0.5y) - 0.4sin(0.8x - 1.3y) + 0.35cos(2x) + 0.2sin(2y)$. A modified 2D line ansatz with 1 qubit and 2 layers was used for the simulation.}
        \label{fig1:subfig1}
    \end{subfigure}
    
    
    \begin{subfigure}[b]{0.67\textwidth}
        \centering
        \includegraphics[width=\textwidth]{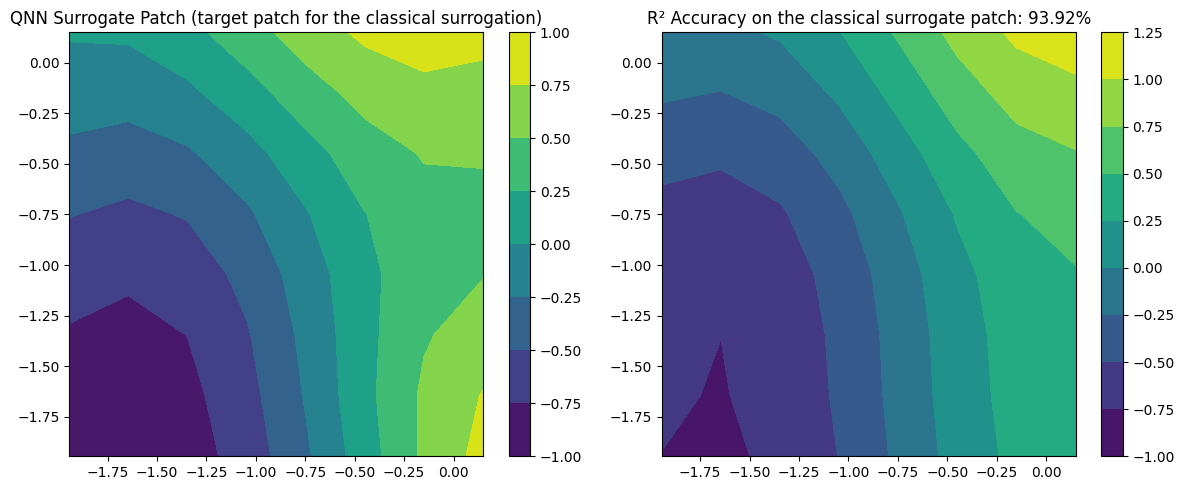}
        \caption{Local classical surrogation of the quantum (QNN) patch sampled from the trained local quantum surrogate, wherein the latter is treated as a black-box model.}
        \label{fig1:subfig2}
    \end{subfigure}

    \caption{Demonstration of local quantum and local classical surrogation.  The $22\times22$ training grid for the full target function on a square domain $x,y\in[-\pi,\pi]$ had a patch of fixed $10\times10$ and $8\times8$ grid size for the local quantum surrogate and the local classical surrogate, respectively.}
    \label{fig1:combined_figures}
\end{figure*}

 In \figref{fig1:subfig1}, we see a demonstration of local quantum surrogation for a fixed patch in the data space of a given target function. We use a modified version of the model described by \cite{PhysRevA.107.062612} wherein a 2-dimensional line ansatz of 1 qubit with 2 layers of reuploading is used for the simulation. As in the original, we too use the Nelson-Mead method for the classical optimization with 500 training and 1500 testing data points for the local region. Thus,
$\text{train\_samples\_per\_axis}
=\Bigl\lfloor\sqrt{500}\Bigr\rfloor
=22$ and
$\text{total grid points}=22\times22 = 484$. $484$ values constitute the full target function
on a square domain $x,y\in[-\pi,\pi]$ in the $22\times22$ training grid. 

The window for the quantum surrogate shown in \figref{fig1:subfig1} is a $10\times10$ grid, i.e., 100 data points.
The immediate observation is that we are able to extract superior performance from these seemingly worse-performing line ansatz models when trained on an appropriately sized local region. 

After creating a local quantum surrogate, we use this surrogate to generate a classical surrogate by implementation of a version of the classical surrogation protocol (wherein the local quantum surrogate is treated as a black-box model) with a patch of $8\times8$ grid size as seen in \figref{fig1:subfig2}. In both surrogation protocols, we observed high $R^2$ values for the local region. The window/patch size for each of the local surrogations here are independent of each other. 


\begin{figure*}[ht]
    \centering
    \begin{subfigure}[b]{0.48\textwidth}
        \centering
        \includegraphics[width=\linewidth]{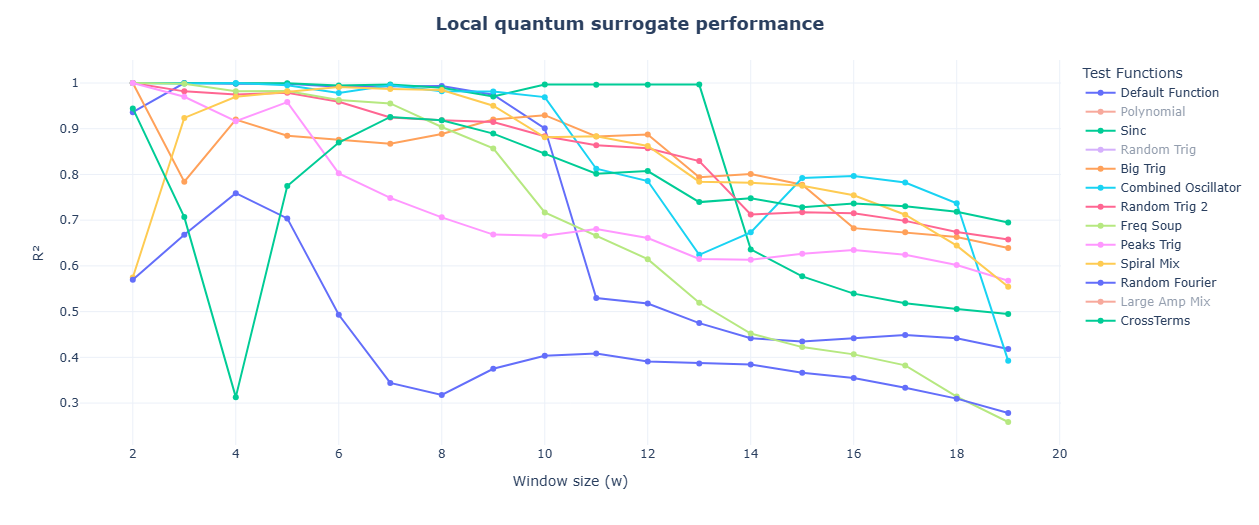}
        \caption{Local quantum surrogate fit to target function (window size vs $R^2$)}  
        \label{fig3:subfig1}
    \end{subfigure}
    \hfill
    \begin{subfigure}[b]{0.48\textwidth}
        \centering
        \includegraphics[width=\linewidth]{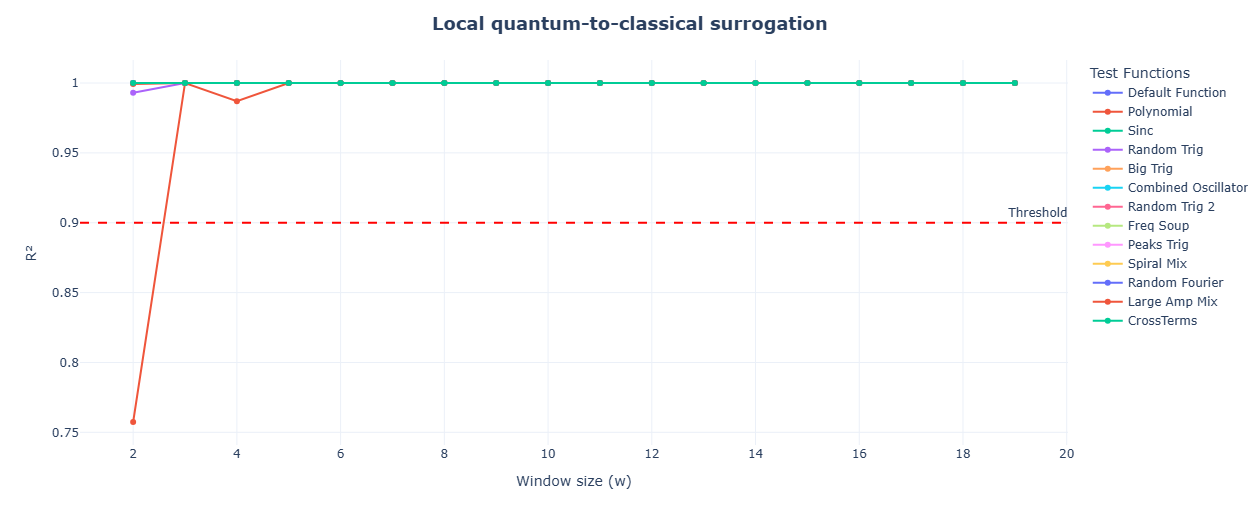}
        \caption{Local classical surrogate fit to the local quantum surrogate, wherein the number of reuploading layers is fed to the local classical surrogate. (window size vs $R^2$)}  
        \label{fig3:subfig3}
    \end{subfigure}

    \vspace{0.5em} 

    \begin{subfigure}[b]{0.48\textwidth}
        \centering
        \includegraphics[width=\linewidth]{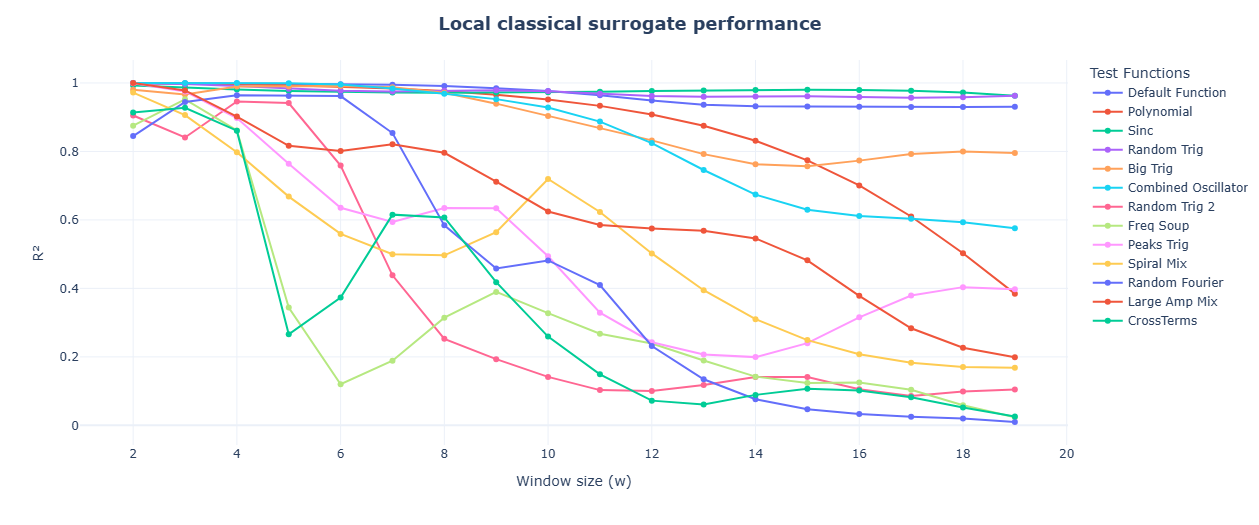}
        \caption{Local classical surrogate fit to target function with the latter treated as a black-box model (window size vs $R^2$)}  
        \label{fig3:subfig2}
    \end{subfigure}
    \hfill
    \begin{subfigure}[b]{0.48\textwidth}
        \centering
        \includegraphics[width=\linewidth]{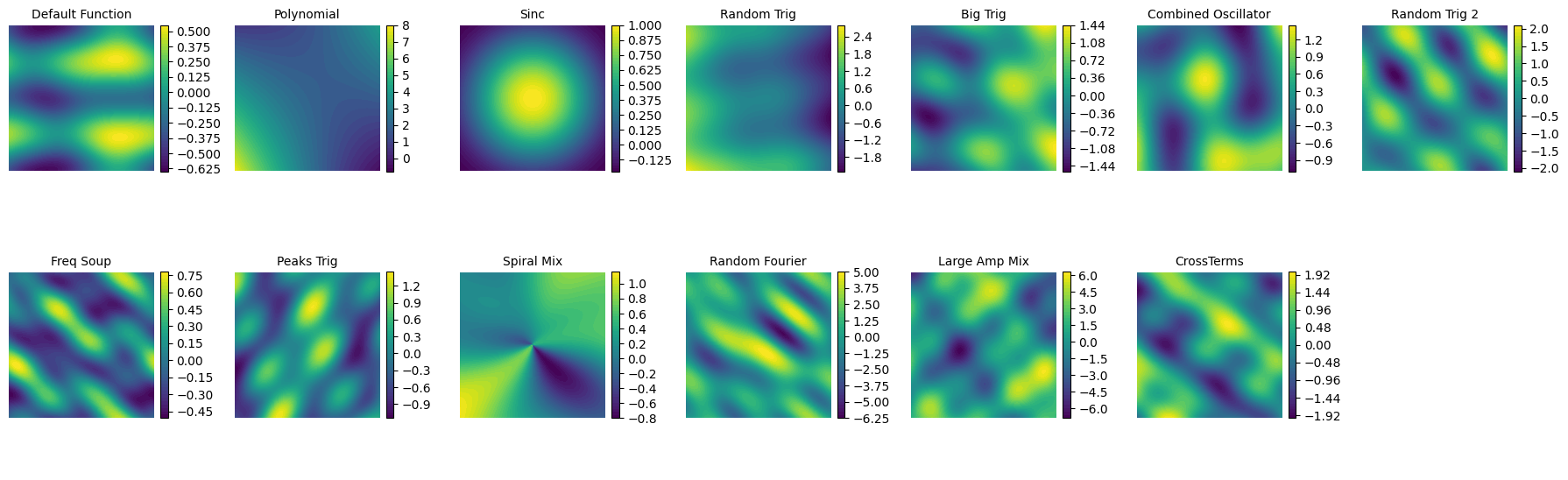}
        \caption{2-D target functions used in the experiment
                 (definitions provided in Table~\ref{tab:smooth}).}
        \label{fig3:subfig4}
    \end{subfigure}

    \caption{Testing the performance of local quantum/classical surrogates in the case of various 2D target functions. Tests had a $22\times22$ training grid for the full target function on a square domain $x,y\in[-\pi,\pi]$.  The simulation looped over different window/patch sizes. The square patch is always anchored at grid index (2, 2) of the $22\times22$ training grid, starting as a $2\times2$ grid and increments in size by 1 grid index in each window cycle until reaching the final size of $20\times20$.}
    \label{fig3:combined_figures}
\end{figure*}

In a similar manner, we trained 1-qubit local quantum surrogates and generated local classical surrogates in the case of various 2D target functions for various window sizes. To stress--test the surrogation pipeline we constructed a suite $\mathcal{F}= {f^{(m)}}_{m=1}^{13}$ of real-valued, twice-differentiable functions $f^{(m)}:[-\pi,\pi]^2\to\mathbb{R}$ (\figref{fig3:subfig4}, Table~\ref{tab:smooth}).  All functions are analytically known, ensuring noise-free ground truth. For 2D signal fitting, we used the same modified line ansatz form of the multidimensional Fourier model and trained on these 13 different 2D target functions. This time, we looped over different window/patch sizes. The square patch is always anchored at grid index (2, 2) (counting from 0) of the $22\times22$ training grid, starting as a $2\times2$ grid and increments in size by one grid index in each window cycle until reaching the size of $20\times20$. Four of these target functions showed significant negative \(R^{2}_{\mathrm{Q}\to\mathrm{T}}\) (local quantum surrogate versus target function) for the local region values and did not converge satisfactorily. We suspect this could be attributed to the limitations of the Fourier-based surrogates. For the rest, we see a clear large \(R^{2}_{\mathrm{Q}\to\mathrm{T}}\) for the local region in the case of small window sizes and a steady drop in learning accuracy as the window sizes increase \figref{fig3:subfig1}, as expected. 

Two Fourier-based classical approximations are implemented: one for the local quantum surrogate and another for the direct classical surrogation on the target functions.

\subsubsection{Local classical surrogate for the local quantum surrogate}

We used the classical surrogation protocol on the local quantum surrogate of depth L generated before and assumed that we always know the number of reuploading layers of the reuploading-based local quantum surrogate. The admissible spectrum is the square
\[
  \Omega_{L}= \bigl\{(k_{1},k_{2})\in\mathbb{Z}^{2}\,:\,
               |k_{1}|,|k_{2}|\le L \bigr\},
  \qquad |\Omega_{L}|=(2L+1)^{2}.
\]

For every window of edge length \(w\) (a \(w\times w\) sub-array of the
22 × 22 training grid, anchored at index (2,2)) we rescale the canonical
\((2L+1)\times(2L+1)\) Cartesian lattice so that all
\((2L+1)^2\) nodes lie inside the patch.  Let's denote those nodes as
\(\mathbf X_{1},\dots,\mathbf X_{(2L+1)^2}\).

We build the design matrix
\[
  \Phi_{ij} = e^{\,i\,\mathbf k_j\cdot\mathbf X_i},
  \qquad \mathbf k_j\in\Omega_L ,
\]
Query the trained quantum surrogate once at each node to obtain  
\(\mathbf y^{\rm q} =
  [f^{\rm q}_{\Theta}(\mathbf X_{1}),\dots,
   f^{\rm q}_{\Theta}(\mathbf X_{(2L+1)^2})]^{\top}\),
and solve the square system \(\Phi\,\boldsymbol\alpha = \mathbf y^{\rm q}\). Exactly $(2L+1)^2$ quantum circuit calls are made.

The resulting local classical surrogate is  
\[
  g_{\boldsymbol\alpha}(\mathbf x)
  = \Re\!\sum_{\mathbf k\in\Omega_L}
        \alpha_{\mathbf k}\,e^{\,i\,\mathbf k\cdot\mathbf x}.
\]


\subsubsection{local classical surrogate for the direct surrogation on the target functions}
The direct surrogate operates on the true target values \(\mathbf{y}^{\mathrm{tar}}\) and we assume that the arbitrary quantum model (predefined target functions in this case) is a blackbox for all our intents and purposes.  We choose a separable spectrum
\(\tilde{\Omega}_{k_{\max}}=\{0\}\cup\{\pm k\}_{k=1}^{k_{\max}}\) with 
\(\,k_{\max}=\lceil w/2\rceil\).  The real–valued basis functions on \(\mathbf{x}=(x_{1},x_{2})\) are
\[
  \Psi(\mathbf{x})
  =\bigl[\,1,\;
    \cos(kx_{1}),\,\sin(kx_{1})_{\,k=1}^{k_{\max}},\;
    \cos(kx_{2}),\,\sin(kx_{2})_{\,k=1}^{k_{\max}}
  \bigr],
\]
giving \(1+4\,k_{\max}\) columns in the matrix \(\Psi_{i j}=\Psi_{j}(\mathbf{X}_{i})\).  The least-squares solution
\[
  \boldsymbol{\beta}
  =(\Psi^{\top}\Psi)^{-1}\,\Psi^{\top}\,\mathbf{y}^{\mathrm{tar}}
\]
minimises the patch MSE, and the separable surrogate is
\[
  g_{\beta}^{\mathrm{sep}}(\mathbf{x})
  =\Psi(\mathbf{x})\,\boldsymbol{\beta}.
\]

For both these local classical surrogations, we compute the coefficient of determination for each window size. For most cases \(R^{2}_{\mathrm{C}\to\mathrm{T}}\) (local classical surrogate versus target function) for the local region of most target functions, we observe an inverse relation between the window size and \(R^{2}_{\mathrm{C}\to\mathrm{T}}\) i.e $R^2$ decreases as window size increases, as seen in \figref{fig3:subfig2}. And for \(R^{2}_{\mathrm{C}\to\mathrm{Q}}\) (local classical surrogate versus local quantum surrogate) for the local region, we observe consistently high (mostly 1 due to the machine precision of the simulations) \(R^{2}_{\mathrm{C}\to\mathrm{Q}}\) values regardless of the window size due to the classical surrogation guarantees associated with the local quantum surrogate based on reuploading in our framework, as seen in \figref{fig3:subfig3}.

\subsection{Demonstration on a QSVM model}
\begin{figure*}[ht]
    \centering
    \begin{subfigure}[b]{\textwidth}
        \centering
        \includegraphics[width=\textwidth]{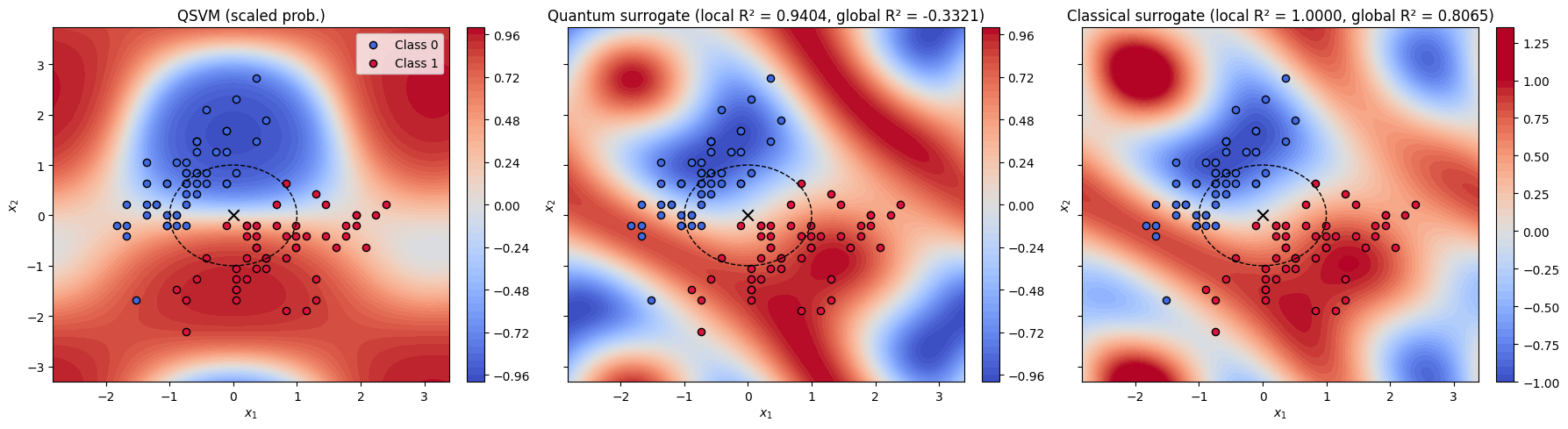}
        \caption{Local and global performances of local quantum surrogate trained on sampled data from the QSVM and local classical surrogate trained on sampled data from local quantum surrogate.}  
        \label{fig4:row1left}
    \end{subfigure}

    \vspace{0.8em}

    \begin{subfigure}[b]{0.67\textwidth}
        \centering
        \includegraphics[width=\textwidth]{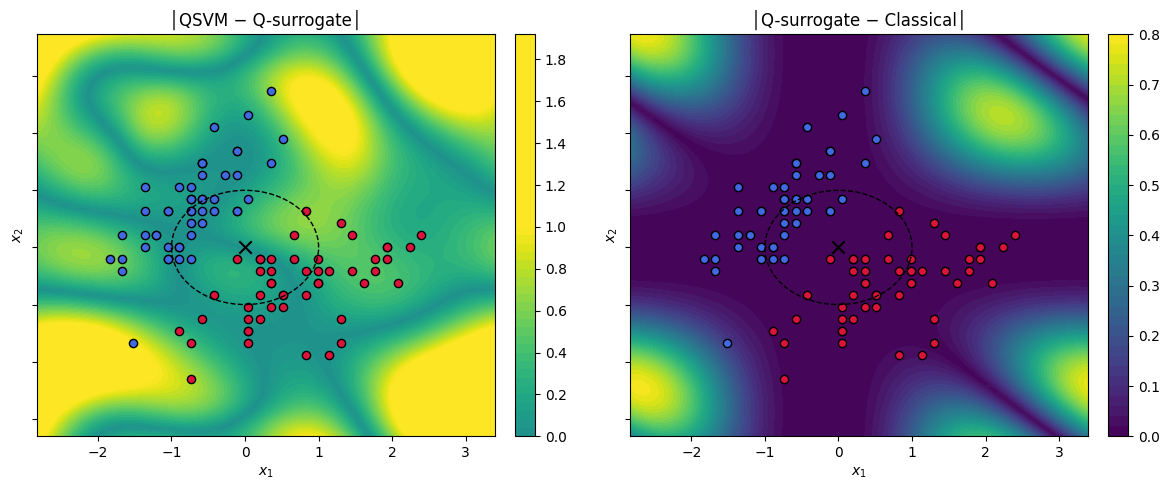}
        \caption{Relative errors of local quantum surrogate against QSVM and local classical surrogate against local quantum surrogate.}  
        \label{fig4:row2left}
    \end{subfigure}

    \caption{Local surrogation protocol on a QSVM trained on a 2-class, 2-feature \textsc{Iris} data set. QSVM had a $R^2$ of 0.84 after its training. PennyLane was used to design the 2-qubit QSVM and the 1-qubit 2-layer reuploading local quantum surrogate. The local quantum-to-classical surrogate is implemented as per the classical surrogation protocol, with the local quantum surrogate treated as a white-box model.}
    \label{fig4:combined}
\end{figure*}

 We now demonstrate our local surrogation
protocol with a quantum support vector machine (QSVM) as our arbitrary quantum model. We used PennyLane \cite{bergholm2022pennylaneautomaticdifferentiationhybrid} to write code for the QSVM and the reuploading-based local quantum surrogate. We adopt the classical \textsc{Iris} data set \cite{iris_53}, $\mathcal{D}_{\mathrm{iris}}=\{(\mathbf{x}_i,y_i)\}_{i=1}^{150}$, 
where $\mathbf{x}_i\in\mathbb{R}^{4}$ encodes sepal and petal statistics and
$y_i\in\{0,1,2\}$ denotes the species label.  
Only the first two classes (setosa and versicolor) and the first two
features (sepal length/width) are retained, yielding
\[
\mathcal{D}=\bigl\{(\mathbf{x}_i,y_i)\in\mathbb{R}^{2}\times\{0,1\}\bigr\}_{i=1}^{100}.
\]
All inputs are standardised by
\(
\tilde{\mathbf{x}}_i=\mathrm{diag}(\sigma)^{-1}(\mathbf{x}_i-\boldsymbol\mu),
\)
where $\boldsymbol\mu$ and $\boldsymbol\sigma$ are the empirical mean and
standard deviation, respectively.
The scaled data occupy the range
\( \tilde{\mathbf{x}}_i\in[-2.6,2.6]^2 \).


A binary QSVM is trained on $\mathcal{D}$ using a data–dependent
\emph{quantum kernel}
\[
K_{\boldsymbol\theta}(\mathbf{x},\mathbf{x}')
\;=\;
\bigl\lvert
\langle 0^{\otimes2}\rvert
U_{\boldsymbol\theta}(\mathbf{x}')^{\dagger}
U_{\boldsymbol\theta}(\mathbf{x})
\lvert 0^{\otimes2}\rangle
\bigr\rvert^{2},
\]
with $U_{\boldsymbol\theta}=U_{\mathrm{ent}}\!
\bigl[R_{Y}(\mathbf{x})\bigr]$ implemented on
two qubits (Pennylane’s \texttt{default.qubit} backend).  Here
$R_Y(\mathbf{x})$ embeds the input angles and
$U_{\mathrm{ent}}$ is a single layer of
nearest–neighbour CNOTs.  The full $100\times100$ kernel matrix is
computed explicitly and supplied to \texttt{SVC} in \textsc{scikit–learn}
with default hyperparameters.
QSVM attains a training accuracy $R^2$ of
$\approx\!85.0\%$. This served as the arbitrary quantum model for our local surrogation protocol, as seen in the first plot of \figref{fig4:row1left}.

Local surrogation is performed on a circular patch
\[
  \mathcal{P}(r)=
  \bigl\{\tilde{\mathbf{x}}_i\in\mathcal{D}\;:\;
         \lVert\tilde{\mathbf{x}}_i\rVert_2<r\bigr\},
  \qquad r>0.
\]
Throughout we fix the radius at \(r=1.0\) centered at (0,0), which selects
\(\lvert\mathcal{P}(1.0)\rvert=30\) points.  
The QSVM posterior
\(p_{\mathrm{Q}}(\mathbf{x})=\Pr\,[y=1\mid \mathbf{x}]\)
is linearly rescaled to \([-1,1]\),
\[
  z_{\mathrm{Q}}(\mathbf{x})=2\,p_{\mathrm{Q}}(\mathbf{x})-1,
\]
and used as a regression target on the patch.

We implemented a 1-qubit local quantum surrogate as per our protocol with PennyLane. A single-qubit feature reupload circuit of depth \(L=2\) is used: 
\[
  U_{\Theta}(\mathbf{x})
  =\prod_{\ell=1}^{2}
     \underbrace{U_{\mathrm{var}}\!\bigl(\bm\theta_{\ell}\bigr)}
                _{\text{StronglyEntangling layer}}\,
     U_{\mathrm{enc}}(\mathbf{x}),
\]
with
\(U_{\mathrm{enc}}(\mathbf{x})=
  U_{\mathrm{ent}}\!
  \bigl[R_Y(x_1),R_Y(x_2)\bigr]U_{\mathrm{ent}}\) and \text{Strongly Entangling layer} template from PennyLane is used to design the trainable block.
The model output is the Pauli-\(Z\) expectation
\(f_{\Theta}(\mathbf{x})=
  \langle0|U_{\Theta}^{\dagger}(\mathbf{x})Z\,U_{\Theta}(\mathbf{x})|0\rangle\). The trainable tensor
\(\Theta\in\mathbb{R}^{2\times 6}\)
(6 Euler angles per upload, 12 angles total)
is initialised
\(\theta_{ij}\sim\mathcal{N}(0,10^{-4})\)
and trained on the patch MSE
\[
  \mathcal{L}_{\mathrm{q}}
    =\frac1{|\mathcal{P}|}
      \sum_{\mathbf{x}\in\mathcal{P}}
      \bigl[f_{\Theta}(\mathbf{x})-z_{\mathrm{Q}}(\mathbf{x})\bigr]^{2},
\]
using 100 steps of Nesterov momentum (\(\eta=0.5\)). For evaluation, a \(100\times100\) Cartesian mesh covering
the bounding box of \(\tilde{\mathcal{D}}\) is generated.
The final surrogate attains
\(
  R^{2}_{\mathrm{Q}\rightarrow\mathrm{T}}(\mathcal{P}) = 0.94.
\) for the local region, as seen in the second plot of \figref{fig4:row1left}.

Local classical surrogation on the local quantum surrogate is performed exactly as outlined before, wherein we assume that the internal parameters of the local quantum surrogate is known. For \(R^{2}_{\mathrm{C}\to\mathrm{Q}}\) (local classical surrogate versus local quantum surrogate) for the local region, we observe consistently high (mostly 1 due to the machine precision of the simulations) \(R^{2}_{\mathrm{C}\to\mathrm{Q}}\) values regardless of the window size due to the classical surrogation guarantees associated with the local quantum surrogate based on reuploading in our framework, as seen in the third plot of \figref{fig4:row1left}.

\begin{figure}[t]            
  \centering

  \begin{subfigure}[b]{\columnwidth}
    \centering
    \includegraphics[width=\linewidth]{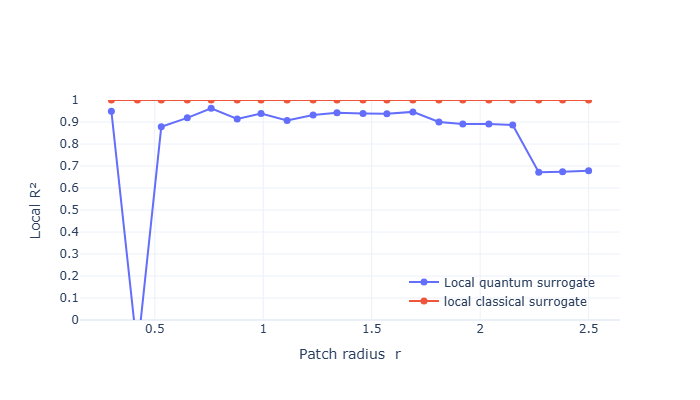}
    \caption{Patch radius \(r\) vs Local \ \(R^{2}\) for the local quantum
             surrogate and its downstream local classical surrogate.}
    \label{fig4:row3left}          
  \end{subfigure}

  \vspace{0.8em}                   

  \begin{subfigure}[b]{\columnwidth}
    \centering
    \includegraphics[width=\linewidth]{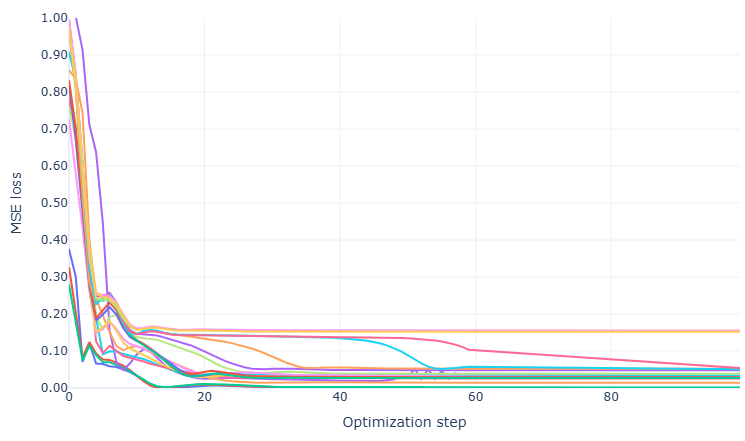}
    \caption{Optimization step vs.\ training loss (MSE) for the same
             set of patch radii in the case of local quantum surrogate; loss converges within few iterations.}
    \label{fig5:stepvsloss}        
  \end{subfigure}

  \caption{Local-surrogate performance (top) and training convergence
           (bottom) trained on the QSVM for varying patch radii from 0.3 radii to 2.5 radii, all centered at \((0,0)\).}
  \label{fig:patch-suite}
\end{figure}

In \figref{fig4:row2left} we see the relative error for both local quantum surrogation and local classical surrogation. We see that within the local region, the relative errors are nominal. In both \figref{fig4:row1left} and \figref{fig4:row2left} we also show the global behavior of the local surrogates. As expected, outside the local region, the accuracy metrics perform poorly for the local quantum surrogates. This is the trade-off we make to achieve high accuracy metrics inside the local region within cost-efficient model constraints. However, for local classical surrogates, the global accuracy metrics perform well given enough data points (although far fewer than the full data space) and given full white-box access to the hyperparameters of the local quantum surrogate. We ran the simulations for various radii patch sizes, starting from 0.3 radii to 2.5 radii, all centered at (0,0) as before. In \figref{fig4:row3left} we plot window size vs $R^2$ and observe the reduction in $R_{local}^2$  for local quantum surrogate as window size increases, local quantum-to-classical surrogation 
 $R_{local}^2$ is 1 regardless of the window size. In \figref{fig5:stepvsloss} we plot the optimization step vs training loss (MSE) for the local quantum surrogate training. Loss converges well in relatively few optimization steps for all these window sizes.

Thus, we have constructed a local quantum surrogate for the QSVM with fewer qubit resources and generated a local classical surrogate for the said local quantum surrogate. This exemplifies the cost reduction as well as the dequantization of the inference phase aspects of the 2-step local surrogation protocol outlined in this paper.

\begin{table*}[h]
  \centering
  \caption{2D target functions used in our experiments}
  \begin{tabular}{@{}l p{0.75\textwidth}@{}}
    \toprule
    \textbf{Function Name} & \textbf{Definition} \\ \midrule

    \texttt{Default Function} &
    $\displaystyle
      -0.02
      +0.04\cos(2x+y)
      +0.25\sin x
      -0.30\cos(2y)
      -0.10\sin(x-y)$ \\[6pt]

    \texttt{Polynomial} &
    $\displaystyle
      0.10\,x^{2}
      -0.20\,y
      +0.30\,xy
      -0.40\,x
      +2.0$ \\[4pt]

    \texttt{Sinc} &
    $\displaystyle
      \frac{\sin r}{r},
      \quad
      r=\sqrt{x^{2}+y^{2}}+\varepsilon$ \\[4pt]

    \texttt{Random Trig} &
    $\displaystyle
      a_{1}\sin x
      +a_{2}\cos y
      +b_{1}\sin(2x+0.5y)
      +b_{2}\cos(2y-0.3x)
      +a_{3}x
      +b_{3}y,
      \;
      a_i,b_i\sim\!U(-1,1)$ \\[8pt]

    \texttt{Big Trig} &
    $\displaystyle
      0.70\sin(0.5x)
      +0.40\sin(2.2y)
      -0.30\cos(2.0x+1.3y)
      +0.25\sin(1.5x-0.8y)
      +0.15\cos(0.9x+2.1y)
      +0.05\sin(3.0x+0.2y)$ \\[10pt]

    \texttt{Combined Oscillator} &
    $\displaystyle
      0.05\,x^{2}-0.10\,y
      +0.60\cos(1.2x+0.5y)
      -0.40\sin(0.8x-1.3y)
      +0.35\cos(2x)
      +0.20\sin(2y)$ \\[10pt]

    \texttt{Random Trig 2} &
    $\displaystyle
      \sum_{i=1}^{3}\!
        \alpha_i\sin(\omega_{x,i}x+\omega_{y,i}y)
      +\beta_i\cos(\tilde\omega_{x,i}x\pm\tilde\omega_{y,i}y),
      \;
      \alpha_i,\beta_i\sim U(-1,1),
      \;
      \omega_{(\cdot)}\in[0.5,2.5]$ \\[10pt]

    \texttt{Freq Soup} &
    $\displaystyle
      0.30\sin(x+2y)
      +0.20\cos(2x+y)
      +0.15\sin(3x+3y)
      +0.10\cos(x-3y)
      +0.05\sin(2x-2y)$ \\[8pt]

    \texttt{Peaks Trig} &
    $\displaystyle
      0.50\cos(1.3x-0.5y)
      +0.40\sin(2.1x+0.7y)
      -0.30\cos(3.0x-2.2y)
      +0.20\sin(1.1x-3.0y)
      +0.10\cos(0.6x+0.3y)$ \\[10pt]

    \texttt{Spiral Mix} &
    $\displaystyle
      0.10\,r
      +0.50\sin(2\theta)
      +0.25\cos(3\theta)
      -0.20\sin(0.5x-0.8y)
      +0.05\,r\cos(5\theta),
      \;
      r=\sqrt{x^{2}+y^{2}},\;
      \theta=\text{atan2}(y,x)$ \\[12pt]

    \texttt{Random Fourier} &
    $\displaystyle
      \sum_{k_x=1}^{K}\sum_{k_y=1}^{K}\!
        \Bigl[
          A_{k_x,k_y}\cos(k_xx+k_yy+\phi_{k_x,k_y})
         +B_{k_x,k_y}\sin(k_xx+k_yy-\phi_{k_x,k_y})
        \Bigr]$ \\[8pt]

    \texttt{Large Amplitude Mix} &
    $\displaystyle
      2.5\sin(1.2x+0.7y)
      -2.0\cos(0.6x-1.1y)
      +1.5\sin(2.0x-2.5y)
      +0.8\cos(3.0x+2.0y)
      -0.3\sin(4.0x-0.2y)$ \\[8pt]

    \texttt{Cross Terms} &
    $\displaystyle
      \sin(x+y)
      +0.5\cos(2x-y)
      -0.4\sin(3y+2.1x)
      +0.3\cos(1.2x-2.5y)
      +0.2\,x\sin y$ \\ 
    \bottomrule
  \end{tabular}
  \label{tab:smooth}
\end{table*}


\subsection{Practical limits of exact classical surrogation protocol}
To demonstrate the exponential blowup in the number of quantum evaluations required with the classical surrogation protocol \cite{Schreiber_2023}, we worked with the \emph{Breast Cancer Wisconsin (Diagnostic)} dataset \cite{breast_cancer_wisconsin_(diagnostic)_17}, containing \(569\) samples with \(30\) real–valued features and
binary labels.  To compress redundancies while retaining \(\approx95\%\) of the total
variance, the feature matrix is projected onto its first \(d=6\) principal
components via PCA.  Each component is then standard-scaled to zero mean and unit
variance, and finally mapped to the interval \([0,2\pi)\) for ease of training with the reuploading model.  The class labels are
mapped from \(\{0,1\}\) to \(\{-1,+1\}\), matching the output range of a Pauli-\(Z\)
expectation value.

Local surrogate training is performed in the scaled PCA space.
For a given hyper-radius \(R\) we retain only those samples  whose absolute value
in \emph{each} coordinate satisfies \(|x_i|<R\); geometrically this is an
axis-aligned hyper-cube of half-width \(R\).  Radii are swept over the grid
\(R\in\{0.5,0.55,\dots ,3.45\}\).  For every hypercube patch we train a
reuploading-type local quantum surrogate \(f_{\boldsymbol\theta}\) with \(n_q=2\) qubits and \(L=3\) data-reuploading
layers in the case of QSVM, written in PennyLane.

\begin{figure*}[ht]
    \centering
    \begin{subfigure}[b]{\textwidth}
        \centering
        \includegraphics[width=\textwidth]{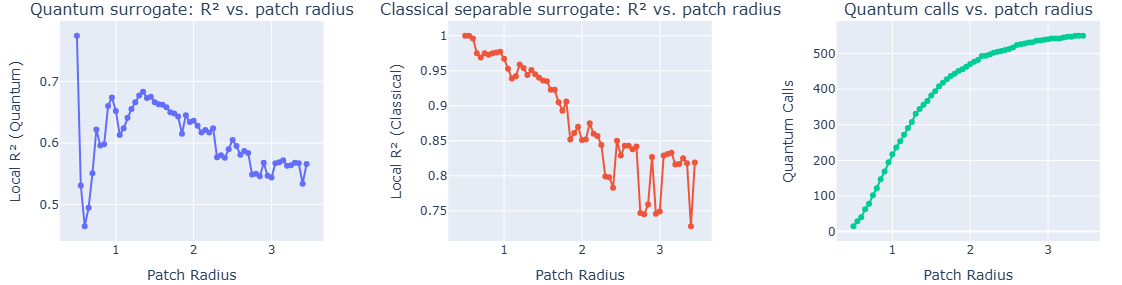}
    \end{subfigure}

    \caption{Local surrogation protocol on PCA reduced 6-feature \textsc{Breast Cancer Wisconsin (Diagnostic)} data set to test the practical limits of classical surrogation protocol. Hypercube patch radii ranged between $0.5-3.5$ with step size $0.05$.  PennyLane was used to design the 2-qubit 3-layer reuploading local quantum surrogate. The local quantum-to-classical surrogate is implemented as per the simplified separable bases Fourier classical surrogate.}
    \label{fig7:combined}
\end{figure*}

A local classical surrogate \(g\) is then fitted \emph{only} to the predictions
\(f_{\boldsymbol\theta}(x)\) on the same patch i.e $n_{local}$. Thus, sample complexity scales linearly with $n_{local}$ under the no-noise assumption.  We now employed a \textbf{separable}
Fourier basis
\[
\Phi(x)=\bigl[1,\;
\cos(kx_j),\sin(kx_j)\bigr]_{k=1,\dots ,L}^{j=1,\dots ,d},
\]
whose design matrix has \(1+2Ld\) (\(37\) columns for \(L=3\) and \(d=6\)).  This lacked the full expressiveness of the Full Fourier basis, but was adequate for this simulation. The least-squares solution \(\alpha=(\Phi^\top\Phi)^{-1}\Phi^\top f_\theta(X)\)
yields \(g(x)=\Phi(x)\alpha\). 
In contrast, the \emph{exact} classical surrogation protocol employs the \emph{full}
tensor-product Fourier lattice \(\exp(i\boldsymbol\omega\!\cdot\!x)\) with
\(\boldsymbol\omega\in\{-L,\dots ,L\}^d\), requiring
\((2L+1)^d\) (\((7^6=117{,}649\) for our particular configuration) basis functions and at least as many samples under the no-noise assumption.
Thus our separable construction slashes the sample complexity by a factor of
\(\approx3\times10^{3}\) while still capturing the dominant single-coordinate
harmonics responsible for the quantum model’s behaviour on each local patch. Due to exponential dependence on the number of features, the exact classical surrogation protocol quickly became computationally expensive for this simulation. In the first and second plots in \figref{fig7:combined} for local quantum and local classical surrogates we see the familiar pattern of reduction in $R^2_{local}$ as the hypercube patch radius increases. Unlike before, we don't see the near-perfect $R^2_{local}$ for classical surrogates. Nevertheless, we observed a $R^2_{local}$ of above 0.7 for our simulations within the given patch range. This is an acceptable tradeoff given the high number of quantum evaluations required with the exact classical surrogation protocol. The third plot shows the quantum calls/evaluations the separable classical surrogate required for each of the patch radii. 

\section{Conclusion}
In this paper, we introduced a local surrogation protocol for quantum learning models. Within the protocol, we introduced reuploading-type based local quantum surrogates. These local quantum surrogates were trained on sampled data within a local region of interest from a fully trained arbitrary quantum learning model. 

Given the classical surrogation guarantees associated with these reuploading-type local quantum surrogates, it is always possible to construct a local classical surrogate for the local region \cite{Schreiber_2023}. This two-step protocol dequantizes and reduces the cost during the inference phase. 
Alternatively, it is possible to directly train a Fourier based local classical surrogate on the sampled data from the arbitrary quantum learning model, but we would lack any tight cost guarantees as in the two-step protocol outlined earlier.

We ran several numerical experiments to validate our two-step protocol, as well as the direct surrogation on the sampled data. In each of these cases, we observed an inverse relation between the size of the local region and the local quantum surrogate's accuracy metrics. This is expected, given that the smaller the size of the local region, the fewer queries we require of the trained arbitrary quantum model, the fewer computational resources, and the easier it is for the Fourier-based local quantum surrogates to fit the sampled data. 

The local classical surrogate constructed from these local quantum surrogates showed near-perfect accuracy metrics as outlined in the classical surrogation protocol. In this paper, the local quantum surrogates served as an intermediary for the construction of the local classical surrogates for the dequantization of the inference phase. But these local quantum surrogates could possibly be used for other purposes, such as in retraining or benchmarking quantum learning models, balancing training between different quantum architectures \cite{bhowmick2025enhancingvariationalquantumalgorithms} and for interpretability purposes, to name a few. An argument can be made that for sufficiently large local quantum surrogates, constructing a classical surrogate may become computationally intractable, even with the classical surrogation protocol, making the local quantum surrogate preferable. Likewise, in the near future, large arbitrary quantum learning models trained on fault-tolerant quantum computers could potentially be surrogated with local quantum surrogates trained on smaller NISQ devices, thus offloading the inference phase onto cheaper, readily accessible hardware while retaining the benefits of fault-tolerant training. Different training schemes, such as sampling the arbitrary quantum learning model on a grid and using that to train the local quantum surrogate could certainly offer high accuracy metrics for a larger range of patch sizes for the same quantum model configurations.

On a similar note, we observed a similar inverse relation between the size of the local region and the directly trained local classical surrogate's accuracy metrics for quite a substantial number of cases. Although we lack any analytical cost guarantees, it might be possible to construct better-performing directly trained local classical surrogates.  As discussed earlier, the exact classical surrogation protocol is well-suited for quantum models trained on low-dimensional data. As there is no explicit dependence on the size of the data space means that for higher-dimensional data, even with localization, this protocol quickly becomes computationally intractable due to the exponentially large number of quantum evaluations it would require. We showed an example of a simpler separable bases Fourier surrogate in such cases, but it works best when the local regions are relatively small. Designing better classical surrogation protocols that achieve near-perfect accuracies without such quantum call blowups is certainly an area for continued research.

We chose the reuploading-type quantum model to construct the framework to deploy our local surrogation protocol as it has analytically proven surrogation guarantees, costs, and a surrogation protocol to back it up. It should be possible to extend the local surrogation protocol to work within other quantum learning architectures as the basis for the local quantum surrogate. One such potential architecture is the matrix product states (MPS) \cite{perezgarcia2007matrixproductstaterepresentations, Bridgeman_2017, PhysRevResearch.6.023218}, but it currently lacks analytical cost guarantees, as seen in the case of reuploading architecture. Therefore, the local surrogation protocol is expected to work effectively when implemented within an appropriate architectural framework.

\section*{Data availability}
All code used in this work, including the ones for generating the figures, is available on the GitHub repository: \url{https://github.com/sreerajrajindrannair/Local-surrogates-for-qml}.

\section*{Acknowledgment}
S.R.N. would like to acknowledge and thank his doctoral research funding agency, Sydney Quantum Academy (SQA), for their continued support. We would also like to thank Afrad Basheer (IQM Quantum Computers), John Azariah (Microsoft, UTS), and Benjamin Southwell (Dolby Labs Sydney) for their valuable discussions on QML and DSP.


\bibliographystyle{ieeetr}
\bibliography{references}

\end{document}